\documentstyle[12pt]{article}
\begin{document}
\newcommand{\id}{{\rm id}}
\newcommand{\diag}{{\rm diag}}
\newcommand{\Hom}{{\rm Hom}}
\newcommand{\rank}{{\rm rank}}
\newcommand{\im}{{\rm im}}
\newcommand{\pr}{{\rm pr}}
\newcommand{\eps}{\epsilon}
\newcommand{\deltaa}{\delta_0}
\newcommand{\rx}{~_R\chi}
\newcommand{\ix}{~_I\chi}
\newcommand{\rh}{~_RH}
\newcommand{\ih}{~_IH}
\newcommand{\ra}{~_RA}
\newcommand{\ia}{~_IA}
\newcommand{\rf}{~_RF}
\newcommand{\If}{~_IF}
\newtheorem{theorem}{Theorem}
\newtheorem{cor}[theorem]{Corollary}
\newtheorem{definition}[theorem]{Definition}
\newtheorem{lemma}[theorem]{Lemma}
\newtheorem{corollary}[theorem]{Corollary}
\newtheorem{prop}[theorem]{Proposition}
\newtheorem{remark}[theorem]{Remark}
\newtheorem{example}[theorem]{Example}
\newtheorem{exercise}[theorem]{Exercise}
\newtheorem{guess}[theorem]{Conjecture}

\title{Topological quantum field theory and crossing number}
\author{{$^{1}$Zhujun Zheng\footnote{Email address: zhengzj@itp.ac.cn}, $^1$Ke 
Wu, $^2$Shikun Wang,  $^{3}$Jianxun Hu}\\
{\small $^1$Institute of Theoretical Physics, Academia Sinica, Beijing 100080, 
P. R. China.}\\        
{\small $^2$CCAST and Institute of Applied Mathematics, Academia Sinica, Beijing 
100080, P. R. China.}\\
                {\small $^3$Department of Mathematics, Zhongshan University, 
Guangzhou, 510275, P. R. China.}}
\date{July 14, 1996}
\maketitle
\begin{center}
\begin{minipage}{120mm}
\vskip 1cm
\begin{center}{\bf Abstract}\end{center}
    {In this paper, we construct a new topological quantum field theory of 
cohomological type and show that 
    its partition function is a crossing number.}
\end{minipage}
\end{center}
\vskip 1cm
\baselineskip 20pt 
\section*{I. Introduction and Notation}
   Topological field theories(TFT), first suggested by M.Atiya[4] based on the 
work of A.Fleer[5], where largely introduced by E. Witten 
   in [1], [3]. It may be grouped into two classes: ``Schwarz type" and 
``cohomological 
   type". In this paper, we will focus on topological field theories of 
cohomological type. Cohomological field theory 
   has many important applications in mathematics and physics. However, the true 
relevance of topological theories to more traditional problems
   in quantum field therory remains to be an open question. It is a very 
instrsting field for many famous mathematicians and physicsts. 

Topological field theories of cohomological type describe intersection theory in 
moduli spaces in the language of local quantum field theory. 
The moduli spaces that arise in physics are very canonical and fundamental 
objects.
In[3] Witten introduced a Lagrangian leading to a topological quantum field 
theory in which the Donaldson invariants of 4-manifolds appear as 
expectation values of the observables. 

It is well know that many topological invariants can be interpreted in terms of 
topological field theories. As an important topological invariant,
crossing number plays an important role in defining Seiberg-Witten invariants 
\cite{sa}.

In this paper, we first  constuct a topological quantum field theory in terms of 
some real fields and some complex fields, then
 develop the path integration  of complex fields. Finally, we show that its 
partition function is the crossing number.

Now we give some notation conventions and some conclusition that will be showed 
in latter sections.

Let $E$ and $F$ be two complex vector bundles over a smooth compact oriented 
manifold with 
real dimension dim$M=2n$. Denote by $E\ominus F$ the equivalent class of a pair 
$(E, F)$ under the equivalence relation 
$(E_1, F_1)\sim (E_2, F_2) \Leftrightarrow E_1\oplus F_2\oplus C^N \cong 
E_2\oplus F_1 \oplus C^N$ for some integer $N$. Let $D$
be section $D\in C^\infty (M, $Hom$(E, F))$ of the bundle of complex linear maps 
$E\rightarrow F$:
$$
\begin{array}{ccccc}
     E &          & \stackrel{D}{\rightarrow} &          & F   \\
       & \searrow &                   & \swarrow &     \\
       &          &        M          &          &
\end{array}.
$$
If
\begin{equation}\label{eq:rank}
      \rank\,E - \rank\,F + n-1 = 0
\end{equation}
then a generic section $D$ will be injective 
at all but finitely many points[2].  
A point $x\in M$ is called a {\bf crossing} if
$
     \ker\,D(x)\ne\{0\}.
$
A crossing is called {\bf regular} if $\dim^c\ker\,D(x)=1$
and
$$
     \im\,D(x) \oplus \left\{(\bigtriangledown_{v}D)(x)\zeta\,|\,v\in 
T_xM\right\}
     = F_x
$$
for some (and hence every) nonzero vector $\zeta\in\ker\,D(x)$.
For every regular crossing define $\nu(x,D)=+1$
or $\nu(x,D)=-1$ according to the orientations in the direct sum,
Then we have

\begin{theorem}\label{theorem:crossing}
Assume~(\ref{eq:rank}) and let
$D\in C^\infty(M,\Hom(E,F))$  be a section with 
only regular crossings.
Then the crossing index of $D$ is given by 
$$
     \nu(D) 
     = \sum_{\ker\,D(x)\ne\{0\}}\nu(x,D)
     = \displaystyle\int_M c_n(F\ominus E).
$$
\end{theorem}

\section*{II.The Topological Quantum Field theory}

In this section, we will construct a cohomological quantum field theory and 
explain why the partition function is the crossing number. 
We will divide the proceedure into three case: Case 1, $E=M\times C;$ Case 2, 
$E=M\times C^{N+1}
$; Case 3, $E$ is a nontrivial bundle.
\vskip 0.8cm
{\bf Case 1.} $E=M\times C$
\vskip 0.8cm
Let $h_{ab}$ be a Hermite metric and $A$ be a  $U(n)$ connection 
on $F$.

Condition (\ref{eq:rank}) says that $\rank F =n$ and a section of $\Hom(E, F)$ 
will be defined by a section $s$ of $F$. A crossing $x$ is 
simply a zero of $s$, it is regular if and only if $s$ intersects the zero 
section of $F$  transverseally at $x$, and the crossing index 
is the intersection number.
 
We will consider a system with a topological symmetry $Q (Q^2=0)$ that carries 
charge one with respect to a  ``ghost number" operator 
$U$. There will be two multiplets. The first consists of local coordinates $u^i$ 
on $M$ (with $U=0$) together with fermions $\psi^i$ 
tangent to $M$, with $U=1$. The transformation laws are
\begin{equation}\label{eq:upsi}
\begin{array}{l}
\delta u^i = i\eps \psi ^i,\\
\delta \psi^i=0,
\end{array}
\end{equation}  
where $\eps$ is an anticommuting parameter. We also define $\delta _0 $ to be 
the variation with $\eps$ removed, for example, 
$\delta_0 u^i=i\psi^i$. The second multiplet consists of an anticommuting 
section $\chi^a$ of $F$ with $U=-1$, and a commuting 
section $H^a$ of $F$; $H$ has $U=0$. The transformation laws are
\begin{equation}\label{eq:chiH}
\begin{array}{l}
\delta \chi^a=\eps H^a - \eps\deltaa u^i A_{ib}^a\chi^b,\\
\delta H^a =\eps\deltaa u^i A_{ib}^a H^b - \frac{\eps}{2}\deltaa u^i 
      \deltaa u^j F_{ijb}^a \chi^b
\end{array}
\end{equation} 
where $F_{ijb}^a$ is the convature of the connection $A$.
The formula  is a covariant version of the more naive $\delta \chi^a=\eps H^a, 
\delta H^a=0.$  

Let 
$\{ e_1, e_2, ..., e_n\}$
be the complex  local frame of $F$. To construct the topological field, we first 
consider $ F$ as a real vector bundle $F^R$. 
Then it is an oriented vector bundle with  orientation $\{e_1, ie_1, e_2, ie_2, 
..., e_n, ie_n\}$. Therefore, we have
\begin{equation}\label{eq:ri}
\begin{array}{l}
\chi=\rx+i\ix, \\
H=\rh+i\ih,\\
A=\ra+i\ia,\\
F=\rf+i\If.
\end{array}
\end{equation}
where $\rx, \ix, \rh, \ih, \ra, \ia, \rf, \If$ are real fields.

Equations(\ref{eq:upsi}), (\ref{eq:chiH}), (\ref{eq:ri} ) imply
\begin{equation}
\begin{array}{l}
\delta \rx^a=\eps \rh^a-\eps\deltaa u^i(\ra_{ib}^a\rx^b - \ia_{ib}^a\ix^b).\\
\delta \ix^a=\eps\ih^a - \eps\deltaa u^i(\ia_{ib}^a\rx^b + \ia_{ib}^a\ix^b),\\
\delta \rh^a=\eps\deltaa u^i(\ra_{ib}^a \rh^b  -\ia_{ib}^a\ih^b) -\frac{\eps} 
         {2}  \deltaa u^i \deltaa u^j(\rf_{ijb}^a \rx^b - \If_{ijb}^a \ix^b),\\
\delta \ih^a= \eps \deltaa u^i (\ra_{ib}^a \ih^b + \ia_{ib}^a \rh^b)
         -\frac{\eps}{2} \deltaa u^i \deltaa u^j (\rf_{ijb}^a \ix^b
         +\If_{ijb}^a \rx^b).
\end{array}
\end{equation}
  
Define
\begin{equation}\label{eq:w}
W=\frac{1}{2\lambda}(\chi, H+2is)
\end{equation}
with $\lambda$ is a small positive real number, 
 the fields $\chi, H, s$ are to be thinked as real fields, $(~,~)$ is the 
Riemannian metric of the bundle $F^R$ induced by the  
 Hermitian metric of $F$.

Define the Lagrangian $L$ to be of the form as follows:
\begin{equation}
L=\delta_0 W
\end{equation}
and 
\begin{equation}
    B=\left(
      \begin{array}{cc}
        \ra_{~b}^a     &     -\ia_{~b}^a \\
        \ia_{~b}^a     &     \ra_{~b}^a
      \end{array}
    \right)
\end{equation}  
\begin{equation}
    G=\left(
      \begin{array}{cc}
        \rf_{~b}^a     &      -\If_{~b}^a \\
        \If_{~b}^a     &      \rf_{~b}^a
       \end{array}
       \right).
\end{equation}    

Since $A$ is an $U(n)$ connection, we have
\begin{equation}
\overline{A}^t=-A  
\end{equation}
Which is equivalalent to the following form,
\begin{equation}
\begin{array}{l}
\ra^t=-\ra,\\
\ia^t=\ia.
\end{array}
\end{equation}
and
\begin{equation}
B^t=-B.
\end{equation}
which imply  that $B$ is an $O(2n)$ connection.

Since
\begin{equation}
\begin{array}{ll}
     F=&dA+A\wedge A\\
     ~=&d\ra + i\ia+ \ra\wedge \ra - \ia\wedge\ia + i(\ia\wedge\ra + 
\ra\wedge\ia),
\end{array}
\end{equation}
we have
\begin{equation}
\begin{array}{c}
dB+B\wedge B
=\left(
\begin{array}{cc}
d\ra     &     -d\ia\\
d\ia     &     d\ra
\end{array}
\right)\\
+
\left(
\begin{array}{cc}
\ra\wedge\ra - \ia\wedge\ia        &     -\ra\wedge\ia - \ia\wedge\ra\\
\ia\wedge\ra + \ra\wedge\ia        &     -\ra\wedge\ia-  \ia\wedge\ra
\end{array}
\right).
\end{array}
\end{equation} 
That is  $G$ is the curvature of the connection $B$.

Now, in view of the above relations, we obtain
\begin{equation}
L=\frac{1}{2\lambda}(H, H+2is)+\frac{1}{4\lambda}(\chi, G_{ijb}^a\psi^i \psi^j 
\chi^b)-\frac{1}{\lambda}(\chi, \bigtriangledown_i^B s \psi ^i ).
\end{equation}

The partition function is defined to be
\begin{equation}
Z=(\frac{\lambda}{2\pi})^n\displaystyle\displaystyle\int {\cal D} u {\cal D} 
\psi {\cal D} \chi {\cal D} H \exp(-L).
\end{equation} 
As a first step to evaluate the above integration , we make an integration over 
$H$ and obtain
\begin{equation}\label{eq:z}
Z=(\frac{\lambda}{2\pi})^n\displaystyle\int {\cal D}u {\cal D}\psi {\cal D} \chi 
\exp(- \frac{(s, s)}{2\lambda}- \frac{1}{\lambda} g_{ab}\chi ^a \frac{\partial 
s^b}{\partial{u^i}} \psi^i+ \frac{1}{2\lambda} 
G_{ijab}\psi^i\psi^j\chi^a\chi^b),
\end{equation}
where $s$ is  transverse to the zero section of $F^R$. Similar to the method 
used in [1], we have 
\begin{equation}
\begin{array}{ll}
  Z=&\lim \limits_{\lambda \rightarrow 0}(\frac{\lambda}{2\pi})^n 
\displaystyle\int {\cal D}u {\cal D}\psi {\cal D} \chi \exp (- \frac{(s, 
s)}{2\lambda}- \frac{1}{\lambda} g_{ab}\chi ^a \frac{\partial 
s^b}{\partial{u^i}} \psi^i+ \frac{1}{2\lambda} 
G_{ijab}\psi^i\psi^j\chi^a\chi^b)\\
  ~=&\nu(D).
\end{array}
\end{equation}

Because the partition function $Z$ is independent of $s$,  it is only necessary 
to consider the case $s=0$, that is  
\begin{equation}
Z=(\frac{\lambda}{2\pi})^n \displaystyle\int {\cal D}u {\cal D}\psi {\cal D} 
\chi \exp (\frac{1}{2\lambda} G_{ijab}\psi^i\psi^j\chi^a\chi^b).
\end{equation}

To evaluate the integration , we firstly choose a proper viebins such that $G$ 
takes the forms
\begin{equation}
\left(
G_{ijab}
\right)=\left(
\begin{array}{ccccc}
0              & \lambda_{ij1} &~~~~~~~~& ~~~~          &~~~\\
-\lambda_{ij1} &   0           &~~~~~~~~& ~~~~          &~~~\\
~~~            & ~~            &\ddots  &~~~~           &~~~\\
~~~            & ~~            & ~~~~~~~& 0             &\lambda_{ijn}\\
 ~~~           &               &        & -\lambda_{ijn}& 0
\end{array}
\right).
\end{equation}

So we have 
\begin{equation}
G_{ijab}\psi^i\psi^j\chi^a\chi^b=\sum\limits_{k=1}^n 2 \lambda_{ijk} \psi^i 
\psi^j \chi^{2k-1} \chi^{2k}.
\end{equation}
The partition function is reduced to 
\begin{equation}
\begin{array}{lll}
Z&=&(\frac{\lambda}{2\pi})^n\ \displaystyle\int {\cal D}u {\cal D}\psi {\cal D} 
\chi \exp (-L)\\
 ~&=&\displaystyle\int {\cal D}u {\cal D}\psi {\cal D} 
\exp(\frac{\lambda_{ijk}\psi^i \psi^j \chi^{2k-1} \chi^{2k}}{\lambda})\\
 ~&=&\displaystyle\int {\cal D}u (\frac{1}{2\pi})^n 
\prod\limits_{k=1}^{n}\lambda_{ijk}.
\end{array}
\end{equation}

In the above viebins, we know that $F$ is diagonal, 
\begin{equation}
F=\left(
\begin{array}{ccc}
i\lambda_1  &    ~   &~   \\
~           & \ddots &~   \\
~           &  ~     &i\lambda_n
\end{array}
\right).
\end{equation}
and can easily calculate the determinant given below
\begin{eqnarray}
\det(\lambda I&-&\frac{i}{2\pi}\diag(i\lambda_1, ..., i\lambda_n))\nonumber\\
&=& \det(\diag(\lambda+\frac{\lambda_1}{2\pi}, ..., 
\lambda+\frac{\lambda_n}{2\pi}))\\
&=&\frac{\lambda_1 \cdots \lambda_n}{(2\pi)^n}+\cdots+\lambda^n \nonumber.
\end{eqnarray}
So we have the Chern class
\begin{equation}
c_n(F)=\frac{1}{(2\pi)^N}\lambda_1\cdots\lambda_n.
\end{equation}
and the crossing number of $D$
\begin{equation}
\begin{array}{lll}
\nu(D)&=&\displaystyle\int_Mc_n(F)\\
   ~  &=&\displaystyle\int_Mc_n(F_E).
\end{array}
\end{equation}
\vskip 0.8cm
{\bf Case 2.}$E=M\times C^{N+1}$
\vskip 0.8cm
Condition (\ref{eq:rank}) says that
$\rank F=n+N.$
Considering that $D$ only have regular crossing, we know that there is a global 
frame $e_1, \cdots, e_N, e_{N+1}$ of $E$ such that
\begin{equation}
\begin{array}{c}
s_1(x)=D(x)e_1,\\
\vdots \\
s_N(x)=D(x)e_N
\end{array}
\end{equation}
is linearly  independent.

Then $\{s_1, \cdots, s_N\}$ span a sunbundle $F_0$ of $F$, and $F_0$ is a 
trivall bundle.

Choose a veibine  $\{f_1, \cdots,f_N, f_{N+1}, \cdots, f_{N+n}\}$ in bundle $F$ 
such that
\begin{equation}
f_i(x)=s_i(x), ~~~~~ 1\leq i \leq N.
\end{equation}
and
\begin{equation}
D=s_1 \oplus \cdots \oplus s_{N+1}.
\end{equation}

In the matrix form, we have
\begin{equation}
D=\left(
\begin{array}{cc}
         I_{N\times N} &     0\\
         0             &    s_{N+1}
\end{array}
\right).
\end{equation}

Let $\overline{s}(x)=[s_N+1]:  M\rightarrow F/F_0.$ Since $D(x)$ only have 
regular crossing, so does $\overline{s},$
and 
\begin{equation}
\nu(D)=\nu(\overline{s}).
\end{equation}

Analogous to the first case, we know
\begin{equation}
\begin{array}{lll}
\nu(\overline{s})&=&\displaystyle\int_M c_n(F/F_0\ominus E)\\
                         &=&\displaystyle\int_M c_n(F).
\end{array}
\end{equation}

\vskip 0.8cm
{\bf Case 3.}$E$ is a nontrivial bundle
\vskip 0.8cm
In the case 1 and 2, we have proved the theorem with $E$ being the trivial 
bundle. However, the general case can easily be reduced to what we have 
disscussed. We know there exist a bundle $E'$ such that $E\oplus E'$ is a 
trivial bundle. Let
\begin{equation}
D'=D\oplus \id: E\oplus E'\rightarrow F\oplus E'.
\end{equation}

$D$ only have regular crossing, so does $D'$. So we have
\begin{eqnarray}
\nu(D)&=&\nu(D') \nonumber\\
      &=&\displaystyle\int_M c_n((F\oplus E')\ominus (E\oplus E'))\\
      &=&\displaystyle\int_M c_n(F\ominus E) .\nonumber
\end{eqnarray}

This prove the theorem.

\end{document}